\documentclass[aps]{revtex4}
\usepackage{epsfig,epsf}
\usepackage{graphicx}
\newcommand{\be}{\begin{equation}}
\newcommand{\ee}{\end{equation}}
\newcommand\beq{\begin{eqnarray}}
\newcommand\eeq{\end{eqnarray}}
\begin{document}

%\emph{Cracow, May 08, 2009}
\begin{flushright}
SAGA-HE-252 \\
\end{flushright}
%\vspace*{3mm}

\title{Andreev reflection between a normal metal and the FFLO superconductor}

\author{Tomasz L. Partyka$^{(1)}$, Mariusz Sadzikowski$^{(1)}$ and Motoi Tachibana$^{(2)}$}

\affiliation{$^{(1)}$ Smoluchowski Institute of Physics, Jagellonian
University, Reymonta 4, 30-059 Krak\'ow, Poland,
$^{(2)}$ Department of Physics, Saga
University, Saga 840-8502, Japan.}

\begin{abstract}
We consider a process of the Andreev reflection between a normal
metal and the s-wave superconductor in the FFLO state. It is shown
that the process takes place if the energy of the incoming electron
is bound within the finite interval called the Andreev window. The
position of the window determines the value of the non-zero total
momentum of Cooper pairs and the value of the gap.
\end{abstract}

%\pacs{74.20.Mn,03.65.Pm,}

\maketitle

\section{Introduction}

During the last two decades one has faced a renaissance of the interest in non-uniform
superconductivity, where the spatial symmetry is broken by a non-zero total momentum $\vec{q}$ of
Cooper pairs \cite{FFLO}. This has happened mainly due to the experimental discoveries of the possible candidates for the
superconductors in the FFLO (Fulde-Ferrel-Larkin-Ovchinnikov) state \cite{exp}. Recently, strong evidence for the existence of the FFLO state has been brought forward for the organic superconductor \cite{org} and for the CeCoIn$_{5}$ superconductor \cite{Ce}. Additionally a
new interest has arisen in the field of strong interaction physics where the new state of matter
called color superconductivity was suggested \cite{col_super}. In condensed matter physics
the FFLO state requires an applied external magnetic field which leads to the Zeeman splitting of the
Fermi surface of conduction electrons, while in QCD the Fermi surfaces of quarks are already split
because of the different masses of different flavor quarks \cite{loff}. It is also a good place to point
out that the non-uniform condensates were also considered in the context of chiral symmetry breaking
\cite{csb_q}, the subject which is still under the current debate \cite{csb_q2}.

The Andreev reflection \cite{andreev} between a normal metal
and an anisotropic superconductor with a directionally dependent gap was discussed in \cite{bruder_hu} with
the main interest concentrated on the d-wave superconductors. In this paper we consider the junction between a normal metal
and the s-wave superconductor in the simplest possible FFLO state where the gap parameter has an oscillating
phase $\Delta (\vec{r})\sim \exp{(i\vec{q}\cdot\vec{r})}$. All the calculations are
performed in the pure Pauli limit. We show that
the Andreev reflection can be used, at least in principle, to detect the existence of the non-zero
total momentum of Cooper pairs. Despite the clear result still a lot of work must be done to bridge between the
ideal case we discuss in this paper and the description of real materials used in experiments.

In the first section we discuss the Bogolubov - de Gennes equations
and the general properties of their solutions. In the second section an exact, numerical
solution is given for some generic parameters. In the last section there is a short discussion
of the possible further work.

\section{Bogolubov - de Gennes equations for the 1-dim LOFF phase}

Bogolubov - de Gennes equation for a non-relativistic superconductor in one dimension takes the form:
\beq
\label{BdG}
Ef(z)&=&-\left(\frac{1}{2m}\frac{d^2}{dz^2}+\mu\right) f(z) + \Delta (z) g(z) \nonumber\\
Eg(z)&=&\left(\frac{1}{2m}\frac{d^2}{dz^2}+\mu\right) g(z) + \Delta^\ast (z)
f(z), \eeq where the FFLO gap function is chosen as $\Delta(z)=\Delta_0e^{2iqz}$ and the FFLO momentum $q$ is assumed
to be non-negative.
The plane wave ansatz $f(z)=f_{k+q}e^{i(k+q)z}$ and $g(z)=g_{k-q}e^{i(k-q)z}$ lead to the matrix equation
\be
\label{BdG_moment}
\left(\begin{array}{cc}
E-\epsilon_{k+q} & \Delta_0 \\
\Delta_0^\ast & E+\epsilon_{k-q}
\end{array}\right)
\left(\begin{array}{l}
f_{k+q} \\
g_{k-q}
\end{array}\right) = 0,
\ee
where $\epsilon_k=k^2/2m-\mu$. The nontrivial solution is possible when
\be
\label{cond}
(E-\epsilon_{k+q})(E+\epsilon_{k-q})-|\Delta_0|^2 = 0
\ee
which gives the dispersion relations
\be
\label{disp_rel}
E_\pm=\frac{1}{2}\left(\epsilon_{k+q}-\epsilon_{k-q}\pm\sqrt{(\epsilon_{k-q}+\epsilon_{k+q})^2+4|\Delta_0|^2}\right) .
\ee
In Fig 1, examples of the dispersion relation (\ref{disp_rel}) are given for
$q=0$ (left panel) and $q\neq 0$ (right panel). It can be immediately seen that the non-zero value of the total
momentum $q$ breaks the parity symmetry $k\rightarrow -k$. The exact shape of the curves depends on the
size of the total momentum $q$. For the values of $0<E_q<\frac{|\Delta_0|^2}{4\mu}$, where $E_q=\frac{q^2}{2m}$,
there are two positive minima of $E_+$ of different depths at the points that are not related by the parity transformation.
The curve $E_-$ resides below zero. However, for $E_q>\frac{|\Delta_0|^2}{4\mu}$ one of the minima of $E_+$ descents below zero.
Simultaneously
a part of the curve $E_-$ emerges above zero on the opposite side of the momentum axis. This phenomenon leads
to more complicated dispersion relations for hole-like and particle-like excitations.

Let us consider the process of the Andreev reflection in a simple
one dimensional geometry configuration where the FFLO superconductor resides at $z>0$
with the gap parameter $\Delta (z)=\Theta (z)\Delta_0 e^{2iqz}$. The normal metal
is at $z<0$ with a junction to superconductor at $z=0$. An electron of the mass $m$ and the energy $E$
above the Fermi energy arrives from the left
at the junction.

\vspace{0.5 cm}
\begin{figure}[h]
\centerline{\epsfxsize=8 cm \epsfbox{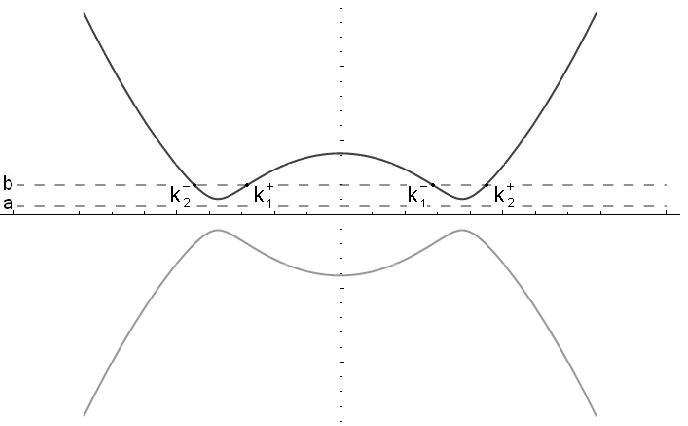} \epsfxsize=8 cm \epsfbox{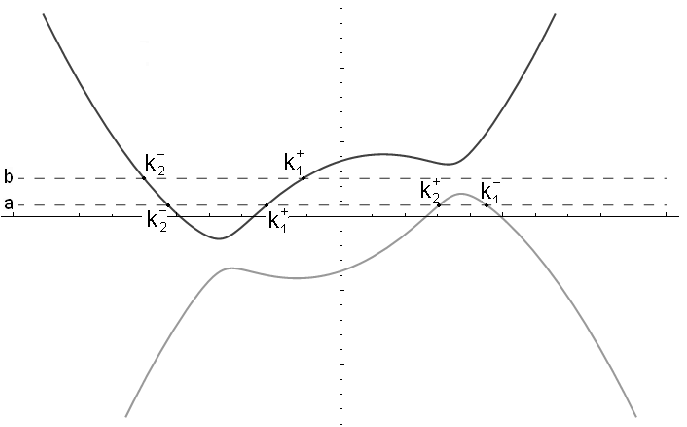}}
\caption{The dispersion relations (\ref{disp_rel}) with the upper curve $E_+$ and the lower
curve $E_-$. In the left panel $q=0$ and in the right panel $q\neq 0$. The lines $a,b$ represents
the energy $E$ of the incoming electron. The crossing points $k_{1,2}^\pm$ describes the solutions
of the equations $E=E_\pm$.}
\end{figure}
\vspace{0.5 cm}

Then one needs to solve the equations (\ref{BdG_moment}) for $f_{k+q}$ and $g_{k-q}$
with the sewing conditions $\psi_<(0)=\psi_>(0), \psi^\prime_<(0)=\psi^\prime_>(0)$ imposed at $z=0$.
The incoming electron of given energy $E$ can excite quasiparticles inside the superconductor
with real or complex momenta. Quasiparticles with the real momenta can propagate inside the superconductor
freely whereas those with the complex momenta penetrate only the region close to the junction.
The solutions with real momenta exist if there are non-zero real solutions of the equations $E_\pm=E$.
Let us first consider the case of normal superconductor with $q=0$.
These solutions geometrically are placed at the crossing of the horizontal lines $a, b$ and the curves $E_\pm$
(lines $a, b$ in the left panel of Fig. 1). For the s-wave BCS superconductor there are four possible real solutions
for $E>\Delta$ (line $b$) among which $k_1^+$ describes hole-like and $k_2^+$ particle-like excitations
propagating from the left to the right whereas $k_1^-,k_2^-$ are those which propagate
in the opposite direction. Only two of these solutions are compatible with the boundary conditions at infinity
(determined by the direction of the incoming electron plane wave). Let us consider the case where the electron
is incoming from the left. Then solutions $k^-_{1,2}$ are rejected and the only left are $k^+_{1,2}$.
It is worth to mention that the incoming electron mainly excites the electron-like quasiparticle
in the superconductor, which follows from the kinematics of the reflection process.
In the case of $E<\Delta$ there are no real solutions (no crossing of the line $a$ with the curve $E_+$
in Fig. 1, left panel). The lack of the freely propagating excitations in the superconductor results in the process
called the Andreev reflection.

In the case of the FFLO superconductor one can immediately infer from Fig. 1 (right panel) that
there are always real solutions of the equations $E_\pm=E$. Let us consider first the solutions
described by the horizontal line $a$. This corresponds to the case of $E>\Delta$ for the uniform BCS superconductor.
There are two solutions $k^+_{1,2}$ of possible four that are compliant with the boundary conditions.
Similarly to the uniform case the electron-like excitation $k^+_{2}$ dominates over the hole-like $k^+_{1}$ quasiparticle.
In the case described by the line $b$ there are two possible solutions $k^+_{1},k_2^-$ and only one solution $k^+_{1}$
is compatible with the boundary condition. This is exactly the one which describes
the hole-like quasiparticle propagating from the left to the right. However, this solution is strongly suppressed by
the kinematics of the reflection.
As a result the quasi-Andreev process takes place when the energy of the
incoming electron is within the gap between the minimum of the upper $E_+$ and the maximum of the lower $E_-$.
Approximately this is in a range
\be
\label{inequal}
\frac{q}{m}\sqrt{2m\mu-q^2}-|\Delta_0|<E<\frac{q}{m}\sqrt{2m\mu-q^2}+|\Delta_0|
\ee
with the corrections of the order of $\frac{q^2}{m}\frac{\Delta}{4\mu}$ to the limits. These inequalities are valid
as long as $E_q<E_F$.

The inequalities (\ref{inequal}) in principle can be used for the determination of the important
superconductor parameters.  The difference between the upper and the lower bound gives
the gap parameter $|\Delta_0|$ whereas the sum determines the value of the total pair momentum $q$
when the Fermi energy $\mu$ and the mass of the charge carrier are given.

\section{Numerical results}

We consider an electron that is injected from the conductor side of the junction. In this case the wave function
takes the form
\be
\psi_{<}(z)=\left(\begin{array}{cc}
1\\
0\\
\end{array} \right) e^{ikz}+B\left(\begin{array}{cc}
1\\
0\\
\end{array} \right)e^{-ikz}+C\left(\begin{array}{cc}
0\\
1\\
\end{array} \right)e^{ipz} ,\;\; z<0,
\ee
\be
\psi_{>}(z)\equiv\left(\begin{array}{cc}
\psi_{\uparrow}\\
\psi_{\downarrow}\\
\end{array} \right) =
F\left(\begin{array}{cc}
f_{1}e^{iqz}\\
g_{1}e^{-iqz}\\
\end{array} \right)e^{ik_{1}^{+}z}+J\left(\begin{array}{cc}
f_{2}e^{iqz}\\
g_{2}e^{-iqz}\\
\end{array} \right)e^{ik_{2}^{+}z} ,\;\; z>0 .
\ee
The total probability current obeys a continuity equation
\begin{eqnarray}
&&\frac{\partial}{\partial t}(|\psi_{\uparrow}|^{2}+|\psi_{\downarrow}|^{2})+ \frac{\partial}{\partial z}\, j_{p} =0,\nonumber\\
j_{p}&=&\frac{1}{2mi}\left(\psi_{\uparrow}^{*}\frac{\partial}{\partial z}\psi_{\downarrow} -
\psi_{\downarrow}\frac{\partial}{\partial z} \psi_{\uparrow}^{*} +
\psi_{\downarrow}^{*}\frac{\partial}{\partial z}\psi_{\uparrow} -\psi_{\uparrow}\frac{\partial}{\partial z}\psi_{\downarrow}^{*} \right) .
\end{eqnarray}
One can decompose the current $j_{p} $ on both sides of the junction into parts
connected with (quasi)hole and (quasi)particle excitations. These are the incident probability current
generated by the incoming electron $j^{<}_{i}=\frac{1}{m}k$,
the probability current connected with the reflected hole $j^{<}_{rh}=-\frac{1}{m} p\left|C\right|^{2}$, the
probability current connected with the reflected electron $j^{<}_{re}=-\frac{1}{m} k\left|B\right|^{2}$. At the
superconducting side one defines the probability current connected with the transmitted quasiparticle
\begin{eqnarray}
j^{>}_{te}&=& \frac{1}{m}\left|J\right|^{2}\left((\left|f_{2}\right|^{2}-\left|g_{2}\right|^{2})\Re[k_{2}^{+}] + q\right) ,
\end{eqnarray}
and the probability current connected with the transmitted quasihole
\begin{eqnarray}
j^{>}_{th}&=&
\frac{1}{m}\left|F\right|^{2}\left((\left|f_{1}\right|^{2}-\left|g_{1}\right|^{2})\Re[k_{1}^{+}] + q\right) .
\end{eqnarray}
Then the hole reflection coefficient is defined as $R_{hole}=|j^{<}_{rh}|/|j^{<}_{i}|$ and the quasiparticle transmission
coefficient is defined as $T_{quasi}=|j^{>}_{te}|/|j^{<}_{i}|$.

\vspace{0.5 cm}
\begin{figure}[h]
\centerline{\epsfxsize=8 cm \epsfbox{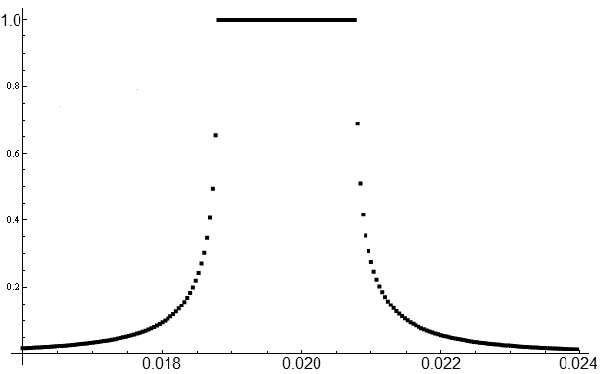} \epsfxsize=8 cm \epsfbox{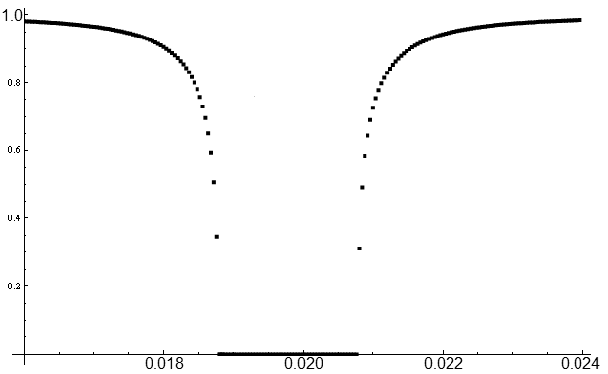}}
\caption{The probability of the hole reflection (left panel)
and the probability of the quasiparticle transition (right panel) as a function of the energy (eV) of the incident particle.}
\end{figure}

For the presentation of our numerical results we set the typical values of the superconducting gap $|\Delta_0| = 0.001$ eV
and the Fermi energy $\mu = 1$ eV. The value of the momentum $q$ corresponds to the wavelength of several interatomic
distances $\sim 10$ nm which gives $q \sim 10$ eV.
The quasi-Andereev reflection, $T_{quasi}\approx$0 and $R_{hole}\approx$1, is expected to occur within the energy range (\ref{inequal}),
that for established parameters induces 0.0187826 eV $\leq$ $E$ $\leq$ 0.0207826 eV. As can be seen in Fig. 2,
this expectation is very well confirmed by the numerical findings. For the incoming electron of the energy outside
the Andreev window (\ref{inequal}) there are solutions of the propagating quasiparticles in the FFLO superconductor,
and they are responsible for the non-zero transmission coefficient $T_{quasi}$.
It is worth to mention that in the opposition to the situation with $q =0$, there are always non-evanescenting
waves in the FFLO. Even within the Andreev window (\ref{inequal}) the solutions with
the real momentum $k$ exists. The amplitude $F$ (responsible for the propagation of quasiholes in the FFLO)
is non zero in this region and the total transmission coefficient $T_{total}=|j^{>}_{te}+j^{>}_{th}|/|j^{<}_{i}|$ is different from zero.
However, the value of $|F|^2$ is negligibly small and is not visible in Fig. 4.

Another interesting quantity is a charge transport on the superconductor side of the junction.
The total charge current obeys the continuity equation \cite{Blonder}
\begin{eqnarray}
e\frac{\partial}{\partial t}(|\psi_{\uparrow}|^{2}-|\psi_{\downarrow}|^{2})+\frac{\partial}{\partial z}\, j^{>}_{c} =
4e\Im[\Delta\psi_{\uparrow}^{*}\psi_{\downarrow}]
\end{eqnarray}
where
\begin{eqnarray}
j^{>}_{c}&=& \frac{e}{m}(\Im[\psi_{\uparrow}^{*}\vec{\nabla}\psi_{\uparrow}] +
\Im[\psi_{\downarrow}^{*}\vec{\nabla}\psi_{\downarrow}]),\\
 \frac{\partial}{\partial z}\,j^{>}_{sc}&=&-4e\Im[\Delta\psi_{\uparrow}^{*}\psi_{\downarrow}] .
\end{eqnarray}
The quasiparticle charge current $j^{>}_{c}$ and the charge current carried by the condensate $j^{>}_{sc}$ take the form:
\begin{eqnarray}
j^{>}_{c} &=&  \frac{e}{m}(\left|F\right|^{2}(\Re[k_{1}^{+}] +
q(\left|f_{1}\right|^{2}-\left|g_{1}\right|^{2})) %\nonumber\\
+\left|J\right|^{2}(\Re[k_{2}^{+}] + q(\left|f_{2}\right|^{2}-\left|g_{2}\right|^{2}))e^{-2\Im(k_{2}^{+})z} \nonumber\\
&+& \Re[Jf_{2}F^{*}f_{1}^{*}(k_{1}^{+*} + k_{2}^{+} + 2q)e^{i(k_{2}^{+}-k_{1}^{+*})z}] %\nonumber\\
+ \Re[Jg_{2}F^{*}g_{1}^{*}(k_{1}^{+*} + k_{2}^{+} - 2q)e^{i(k_{2}^{+}-k_{1}^{+*})z}]) ,\\
j^{>}_{sc} &=&  4e\Delta_{0}\Im[
\frac{\left|J\right|^{2}f_{2}^{*}g_{2}}{2\Im(k_{2}^{+})}{(1-e^{-2\Im(k_{2}^{+})z})} %\nonumber\\
+ i\frac{J^{*}f_{2}^{*}Fg_{1}}{(k_{1}^{+} - k_{2}^{+*})}{(1-e^{i(k_{1}^{+} - k_{2}^{+*})z})} \nonumber\\
&+& i\frac{F^{*}f_{1}^{*}Jg_{2}}{(k_{2}^{+} - k_{1}^{+*})}{(1-e^{i(k_{2}^{+} - k_{1}^{+*})z})}] .
\label{charge-current}
\end{eqnarray}
The main contribution to the charge current originates from the part of the equation (\ref{charge-current}) that is proportional to
the coefficient $|J|^{2}$, that is:
\begin{eqnarray}
j^{>}_{sc}\approx \frac{2e\Delta_{0}\left|J\right|^{2}\Im(f_{2}^{*}g_{2})}{\Im(k_{2}^{+})}\left(1-e^{-2\Im(k_{2}^{+})z}\right) .
\label{approx-current}
\end{eqnarray}
Considering the limit $q\rightarrow 0$ the charge current has a simple form
\begin{eqnarray}
j^{>}_{sc}(q=0)=2ev_{F}\left(1-e^{-\frac{2\sqrt{|\Delta|^{2}-E^{2}}}{v_{F}}z}\right)
\end{eqnarray}
where $v_{F}=\sqrt{\frac{2\mu}{m}}$  is the Fermi velocity. For $q\neq 0$ one can rewrite equation (\ref{approx-current})
in the approximate form:
\begin{eqnarray}
j^{>}_{sc}\approx2ev_{F}\left[1-\exp\left(-\frac{2\sqrt{|\Delta|^{2}-
(E-\frac{q}{m}\sqrt{2m\mu-q^{2}})^{2}}}{\sqrt{v_{F}^{2}-\frac{q^{2}}{m^{2}}}}z\right)\right] .
\end{eqnarray}
Here one introduces a standard parameter $\xi$ that describes the penetration depth of the evanescenting currents
\begin{eqnarray}\xi(q=0)=\frac{v_{F}}{2\sqrt{|\Delta|^{2}-E^{2}}},\quad
\xi(q)=\frac{\sqrt{v_{F}^{2}-\frac{q^{2}}{m^{2}}}}{2\sqrt{|\Delta|^{2}-(E-\frac{q}{m}\sqrt{2m\mu-q^{2}})^{2}}} .
\end{eqnarray}
\begin{figure}[t]
\centerline{\epsfxsize=8 cm \epsfbox{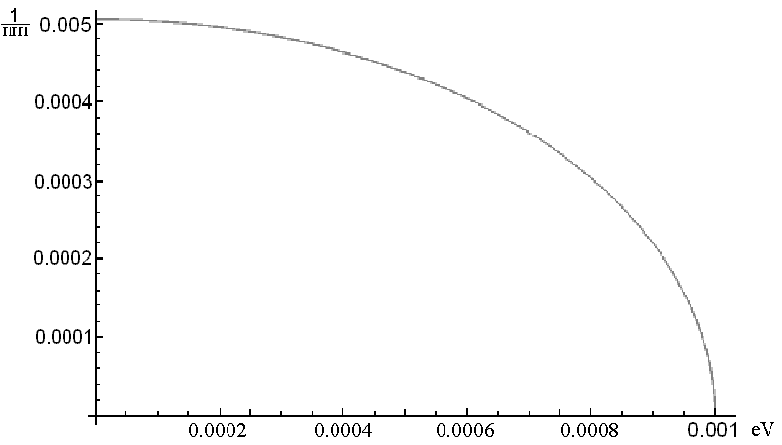}\epsfxsize=8 cm \epsfbox{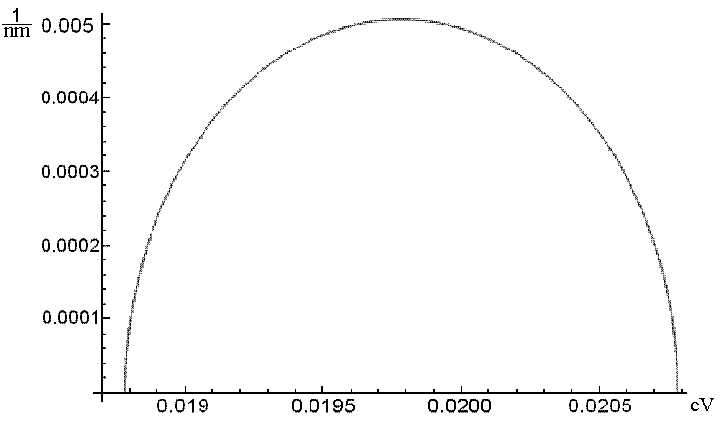}}
\caption{The plot of $1/\xi$ $[nm^{-1}]$  as a function of the energy [eV] of the incoming particle. Left chart q=0, right chart q=10 eV.}
\end{figure}
\begin{figure}[h]
  \begin{center}
  \includegraphics[width=0.75\textwidth]{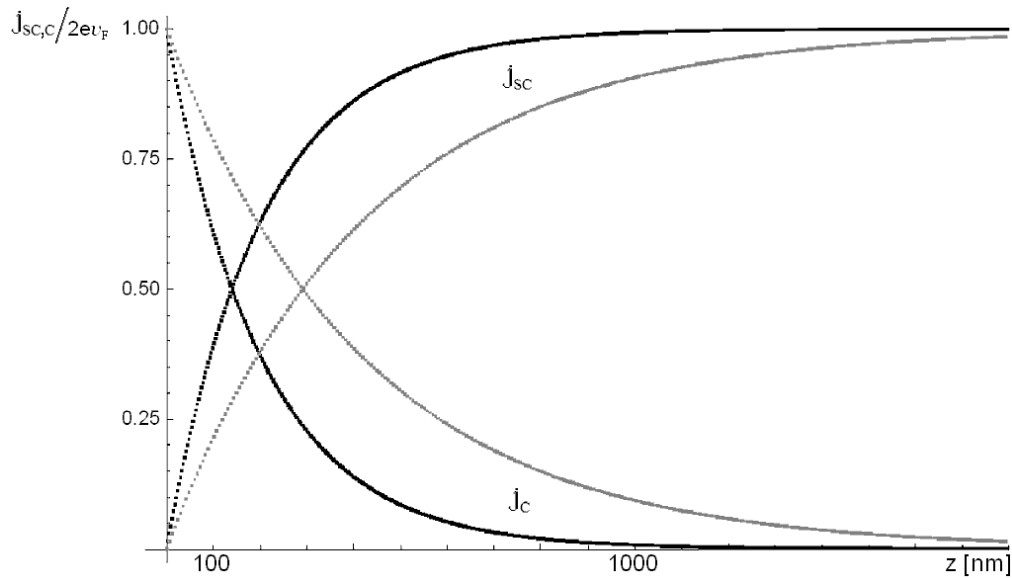}
\end{center}
\caption{The charge current and the current carried by the condensate in a units of $2ev_{F}$ 
as a function of a distance $z$ [nm] from the junction at $z=0$ (black points refers to the incident
energy $E= 0.02$ eV, gray points to energy $E= 0.0189$ eV ).}
\end{figure}
In Fig. 3 we present the penetration depth $\xi $ as a function of energy of the incoming electron.
We keep the values of $m$, $\Delta$ and $\mu$ constant. As one could expect, when $q\neq0$ penetration
distance approaches towards infinity as $E$ approaches towards the lower or higher limit at the Andreev window (\ref{inequal}).
The minimum value is at $E=\frac{q}{m}\sqrt{2m\mu-q^{2}}$.
On the contrary, for $q=0$, the maximum penetration distance is at $E=\Delta$ and the minimum is reached at zero energy.
The dependence of $\xi$ on the energy is also seen in Fig. 4, that presents the conversion of the normal charge current
into the supercurrent at the FFLO phase.

\section{Conclusions}

In this paper we have considered a process of the Andreev reflection
between a normal metal and s-wave superconductor in the FFLO state
with a single plane wave in the pure Pauli limit. We have found that the
Andreev process takes place only for electrons with the energy
located within the Andreev window given by equation (\ref{inequal}).
This is clearly visible in the dependence of transmission and
reflection coefficients associated with the energy of the incoming electrons. The
other interesting quantities are the charge currents flowing through
the junction and the penetration depth of the excitation inside
superconductor. This last parameter depends on the energy of the
incoming electron in very different way compared to the s-wave BCS
superconductor (Fig. 3). In conclusion the Andreev process, in
principle, can serve as a good probe to look for the FFLO state.
Obviously the model given here is a far reaching idealization and
additional works must be done to bring the results closer to
experimental conditions. However, the results presented in this
paper are robust in a sense that they are the simple consequences of
the existence of the non-zero total momentum of Cooper pairs.

The next interesting step is analysis of the FFLO state with an
oscillating value of the gap parameter $\Delta\sim
\cos(\vec{q}\cdot\vec{x})$. Another important point is to adopt the
external magnetic field in the picture and try to go off the pure
Pauli limit. One can also extend the analysis of the Andreev
reflection in the color superconductors \cite{andreev_sad} for the
FFLO states.

\textbf{Acknowledgement:} We would like to thank Taku Waseda for
joining us in the very early stage of this work. The research of
M.S. is supported in part by the MEiN grant N N202 128736 (2009 -
2012). M.S. would like to thank the Saga University for hospitality. 
We also would like to thank Professor Jozef Spalek for many interesting discussions.

\end{document}